# Autonomous Transition State Search with Soft Actor-Critic Reinforcement Learning

Utham Suresh, Konstantinos D. Vogiatzis*

*Department of Chemistry, University of Tennessee, Knoxville, TN, 37996-1600, The United States*

*Corresponding Author: kvogiatz@utk.edu

## Abstract

Transition state (TS) search is a crucial step in understanding chemical reactivity and mechanisms, yet conventional algorithms remain computationally intensive and heavily reliant on initial guesses, user's expertise, and chemical intuition. While recent machine learning approaches have shown promise, they demand either large training datasets or geometric interpolation between known endpoints, limiting their generality. In this work, we introduce a TS search model based on the soft actor-critic model, an advanced reinforcement learning algorithm in which an agent learns to navigate potential energy surfaces directly from local energetic and curvature information starting from a given reactant and its corresponding product. By formulating the search as a sequential decision-making process in internal coordinates, the agent adaptively proposes chemically meaningful structural updates through a reward function designed to promote movement towards saddle point regions. Without labelled trajectories or prescribed reaction pathways, the method successfully identifies TS geometries for standard benchmark reactions, operating directly on realistic molecular potential energy surfaces. These results highlight the potential of RL as a general strategy for reducing dependence on initial guesses and enabling scalable, automated reaction discovery across diverse chemical systems.

Table of Contents Graphic

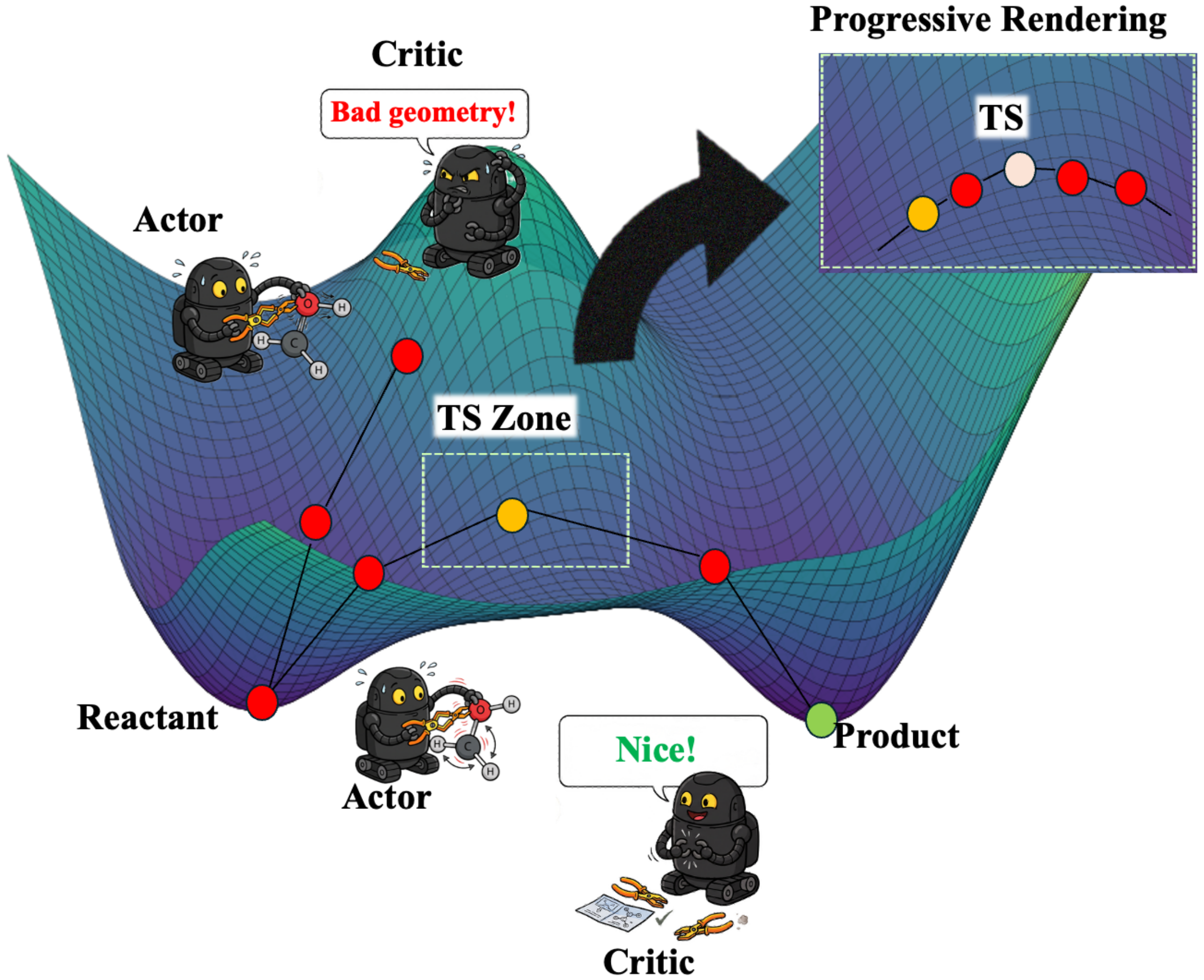

## Introduction

In recent years, the rapid computational mapping of chemical reaction networks has become increasingly important for accelerating molecular discovery,[1, 2] enabling the design of sustainable catalysts[3] and the identification of promising reaction pathways.[4-7] Central to this effort is the accurate identification of transition states (TSs), which determine the kinetic feasibility of chemical reactions[8, 9] and are inherently difficult to characterize experimentally because of their transient nature.[10, 11] Quantum chemical methods based on the first and second derivatives of energy with respect to atomic coordinates have been widely used to accurately locate saddle points on potential energy surfaces (PESs).[12] Single-ended approaches such as the Berny algorithm,[13] the artificial force induced reaction (AFIR) method,[14] and the anharmonic downward distortion following (ADDF) approach[15] can reliably converge to a TS when supplied with a suitable initial geometry. Their efficiency and success rate, however, remain strongly dependent on the quality of this initial guess, which in practice often relies on expert chemical intuition limiting both automation and scalability. Double-ended methods, notably the nudged elastic band (NEB)[16] and the growing string methods (GSMs),[17-19] address the initial-guess problem by interpolating a minimum energy pathway between known reactant and product geometries and identifying the highest-energy point along this path as the TS candidate. Although these approaches are widely adopted in practice, they are computationally expensive. They require many intermediate images to be evaluated at the quantum chemical level and can struggle to converge when reactions involve large structural changes or competing pathways.

The growing adoption of machine learning (ML) in chemical research has introduced complementary strategies for accelerating TS searches,[20-28] ranging from machine-learning interatomic potentials (MLIPs)[29-31] that approximate the potential energy surface to reduce the cost of physics-based optimization algorithms, to generative models[32-36] capable of directly predicting saddle-point geometries without exhaustive surface exploration. While these approaches have significantly reduced computational overhead, many depend on large, precomputed datasets for training, a requirement that constrains their transferability across chemically diverse systems and limits their applicability to reaction classes not represented in the training data.

Reinforcement learning (RL)[37] offers a fundamentally different paradigm for locating critical points on a PES. Unlike supervised learning which relies on pre-existing datasets, RL agents dynamically interact with the environment,[38, 39] observe the results of their actions and adapt based on their feedback as the process unfolds. This process allows the agents to solve real-time problems without requiring prior knowledge of the environment. By treating the PES as a dynamic environment,[40-42] RL agents have demonstrated their ability to discover optimal reaction pathways through active exploration with the help of high-order geometric information such as gradients and curvatures. Among the earliest applications of RL to PES exploration, Xu et al.[43] employed Q-learning on precomputed energy surfaces that were discretized along user-defined reaction coordinates. This coarse-to-fine strategy using low-level calculations to identify promising regions before selectively refining them at higher accuracy, is effective for smaller systems as they rely on precomputed PESs. Barrett and Westermayr[44] extended these ideas beyond grid navigation by integrating an actor-critic agent into the NEB framework in which a PaiNN equivariant network[45] learned atomic displacements through combinations of NEB spring and potential forces. The method successfully located TSs for Claisen rearrangement and $S_N2$ reactions, although its reliance on pretrained machine-learning potentials and the NEB force framework may limit broader applicability. Recently, the implementation of transfer learning from pre-trained agents on foundational energy landscapes has demonstrated the potential to accelerate TS discovery in previously unseen systems, suggesting that RL-based strategies offer an efficient route for navigating complex molecular potential energy surfaces.[46]

In this work, we present the soft actor-critic transition state search (SACTS) model, an agent built for autonomous TS discovery that operates without reliance on human chemical intuition or precomputed reaction path data. SACTS interacts directly with a semiempirical quantum chemical environment, utilizing energy gradients and Hessian eigenvalues as observational feedback to navigate the PES and identify first-order saddle points through a structured curriculum that transitions from broad surface exploration to fine convergence in the vicinity of the TS. To the best of our knowledge, SACTS represents the first reinforcement learning framework deployed for direct PES exploration on real molecular systems for the purpose of TS discovery with absolute control over all the internal coordinates of the molecule and no prior computed pathway or human intuition. The framework is benchmarked across a diverse set of chemical reactions, demonstrating its capacity to locate TSs without system-specific initialization or expert guidance. The obtained results establish SACTS as a viable and efficient strategy for automated TS discovery, offering a transferable and data-light alternative to existing machine learning approaches.

# Methods

## Discrete Grid Reinforcement Learning

At its core, a RL algorithm involves an agent that interacts with an environment to maximize cumulative rewards.[47] Through repeated trial-and-error interactions, the agent gathers experience that is used to learn an optimal decision-making strategy, often represented as a policy $\pi$. As this policy improves with accumulated experience, the agent becomes increasingly effective at selecting actions that lead to successful task completion. This learning process is commonly formulated as a Markov Decision Process (MDP), which provides a simple yet powerful framework for modelling sequential decision-making in stochastic environments.[48, 49] An MDP consists of states, actions, transition probabilities, and rewards. The states $s$, satisfying the Markov property, represent the information describing the environment at a given time, while the actions $a$ enable transitions between states according to the transition probability function $P(s' \mid s, a)$. The reward function $R(s, a)$ evaluates the desirability of taking a particular action in a given state. Within this setting, the overarching objective is to maximize the cumulative reward or total return expressed as:

$$J(\pi) = \mathbb{E}_{\pi}\left[\sum_{t=0}^{\infty} \gamma^{t} R(s_t, a_t)\right], \tag{1}$$

Where $\gamma \in [0,1]$ is the discount factor that controls the trade-off between immediate and future reward.[50] To formulate transition-state search within the RL framework, the potential energy surface (PES) is defined as the agent's environment with the individual molecular geometries as its states. Transitions between these geometries are facilitated by actions, defined as structural modifications or coordinate shifts intended to guide the agent toward the target saddle point.

To provide the agent with a sufficiently descriptive understanding of the PES, various representation strategies have been explored in the literature. Discrete grid-based mapping[43] employs a state representation where the high-dimensional PES is coarse-grained into a grid-like maze, where each cell representing a specific region of the configuration space. In this model, actions are defined as discrete moves between adjacent cells. This allows the agent to navigate the surface by selecting one of several possible directions at each step, with rewards based on the energy height of the traversed cells to identify the minimum energy pathway. Crucially, this approach incorporates a coarse-to-fine refinement strategy in which, once a candidate TS region is localized at a low resolution, the algorithm progressively increases the grid density around that point. This iterative narrowing of the configuration space enables the agent to refine the coarse geometry into a precise saddle-point structure, effectively balancing global exploration with local accuracy. While effective for systems that can be described by a small number of reaction coordinates, the grid-based formulation restricts the representation to the degrees of freedom explicitly selected for the scan, meaning that the remaining atoms in the molecule are held fixed and cannot adjust in response to the evolving reaction coordinate. For larger molecules where spectator groups undergo subtle geometric relaxation during the reaction, this frozen environment approximation may introduce systematic error into the predicted saddle point geometry. The discrete nature of the state and action spaces also imposes practical constraints on scalability, as the number of grid cells and associated stored values grows combinatorially with the number of coordinates included in the representation. Alternatively, curvature-informed continuous exploration[46] makes use of higher-order geometric information from the PES to construct a continuous state representation, by implementing multi-step history of gradients, internal coordinates and PES curvatures (eigenvalues) as a state vector. Actions in these systems are defined as continuous shifts in internal coordinates, aligned with the softest curvature of the PES to facilitate the climb toward a saddle point followed by the presence of a reward function that depends on the characteristic feature of a saddle-point which is, the presence of a single imaginary frequency with near-zero gradient. Here, we have adopted this continuous representation strategy, as it naturally accommodates the full set of molecular degrees of freedom without requiring prior selection of reaction coordinates with the specific state features, action space, and reward structure described in the following sections.

## The Soft Actor-Critic Transition State Search (SACTS) Model

Extending these established methodologies, we have developed the soft actor-critic transition state Search (SACTS) model, an agent designed to autonomously explore the PES and locate accurate TS geometries. Starting from the reactant geometry, the agent navigates the PES toward the product to localize first-order saddle points. To ensure chemical efficiency, all geometric communications and structural modifications are performed in the Z-matrix format, with the geomeTRIC package[51] facilitating seamless transformations between Cartesian and internal coordinate spaces. The search process is divided into distinct phases, allowing the agent to transition from broad global exploration to high-precision localization. Within this architecture, the design of the state representation, action space, and reward function is central to enabling the agent to learn generalized physical intuitions. By prioritizing these physically grounded descriptors over system-specific parameters, the SACTS framework functions as a robust, autonomous model capable of navigating diverse and unknown PESs.

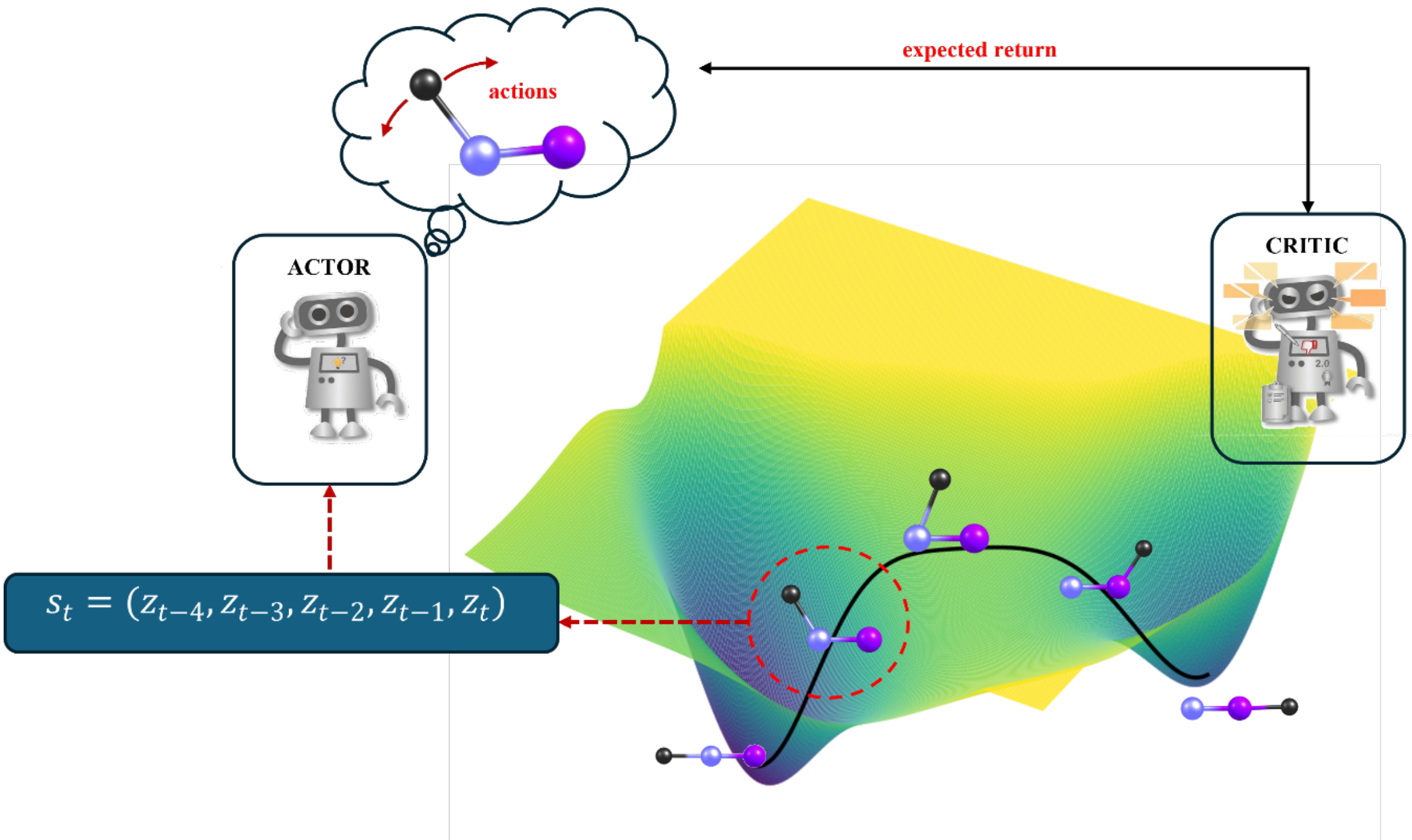

**Figure 1.** Schematic overview of the soft actor-critic transition state Search (SACTS) model. The actor proposes geometric displacements of the molecular structure (actions) based on the stacked-history state vector $s_t$, while the critic evaluates the expected return to guide the actor toward first-order saddle points on the potential energy surface.

## State Representation

The effectiveness of any reinforcement-based exploration depends strongly on the choice of state representation. For transition-state search, gradient and curvature information provide a natural and physically meaningful description of the local topology of the PES. Accordingly, these descriptors, along with additional structural and safety constraints, form the state representation for our agent.

The first group of features captures the topological character of the current point on the surface. The log-transformed gradient norm ($g_{\text{norm}}$) normalizes force magnitudes across different regions that the agent explores, while the eigenvalues of the Hessian are compressed through a $tanh$ filter to preserve critical sign changes during saddle-point localization ($\lambda_{\text{soft}}$). An explicit imaginary mode count ($n_{\text{neg}}$) complements these continuous descriptors, and a temporal gradient delta ($\Delta g_{\text{norm}}$) computed between successive steps helps the agent to anticipate stationary points. A second group of features provides directional guidance relative to the target product. This includes the dot product between the softest vibrational mode and the target direction vector ($v_{\text{targ}}$), as well as the distance to the product in terms of internal coordinates ($d_{\text{target}}$). A slope sensor ($v_{\text{grad}}$) further identifies whether the trajectory is proceeding uphill or downhill along the reaction coordinate. Finally, to ensure that the agent remains within a chemically reasonable configuration space, a repulsive proximity signal is introduced as a safety feature. This signal is computed from the minimum distance between non-bonded atoms, as defined by internal coordinates,

$$s_{\text{prox}} = \left[\frac{0.74}{r}\right]^6 \tag{2}$$

and grows sharply as atoms approach unphysically close. The eight features described above are concatenated into a single observation vector $z_t$,

$$z_t = \left(g_{norm},, n_{neg}, \Delta g_{norm}, v_{targ}, v_{grad}, d_{target}, s_{prox}, \lambda_{soft}\right). \tag{3}$$

Following the approach of Ahuja et al.[40], we stack the current observation with the four preceding steps to construct the full state representation,

$$s_t = (z_{t-4}, z_{t-3}, z_{t-2}, z_{t-1}, z_t). \tag{4}$$

This temporal context state is essential for distinguishing between a random fluctuation and a sustained climb toward the TS, providing the agent with short-term memory to capture the dynamics of PES traversal. In the event of an electronic structure failure or detection of an unphysical geometry, the episode is terminated immediately, and the agent receives a penalty reward. The returned observation in such cases is the last valid stacked history state, ensuring that erroneous geometries do not corrupt the feature buffer used for policy gradient estimation.

## Action Representation

The agent operates within a continuous three-dimensional action space, $\mathbf{a} \in [-1,1]^3$, wherein each dimension governs a geometrically distinct contribution to the displacement taken on the potential energy surface at each timestep. To construct a meaningful and chemically interpretable step, the local PES topology is first decomposed into two orthogonal basis directions at every point along the trajectory. The primary direction, termed the soft mode $\hat{v}_{\text{soft}}$, is defined as the eigenvector of the internal-coordinate Hessian $\mathrm{H}_q$ corresponding to its algebraically smallest eigenvalue, $\lambda_{\min}$ in the direction of the reaction, and its sign is resolved at each step by enforcing a positive projection onto the instantaneous reactant-to-product displacement vector $\hat{d}_{\text{target}}$. This displacement vector is computed from the full set of internal coordinates for all reactions studied in this work. Additionally, a topology-aware variant was placed for the ring opening reaction, in which the target vector is masked to zero out Z-matrix entries belonging to spectator atoms that do not participate in bond breaking, as the large collective rearrangement in such reactions can dilute the directional signal toward the chemically active coordinates. Both the masked and unmasked formulations for this system are compared in the results section. The gradient $\mathbf{g}_q$, expressed in internal coordinates, is subsequently decomposed into components parallel and perpendicular to $\hat{v}_{\text{soft}}$, yielding a steering direction:

$$\hat{v}_{\text{steer}} = \frac{\mathbf{g}_\perp}{\|\mathbf{g}_\perp\|} \tag{5}$$

With these two vectors established, the composite internal-coordinate step is assembled as:

$$\Delta q_{raw} = a_1\, \hat{v}_{\text{soft}} + a_2\, \hat{v}_{\text{steer}} \tag{6}$$

where, these two actions of the agent serve as learned scaling coefficients for each of the two basis directions. A third action $a_3$ (Eq. 7) operates as a global step-size multiplier and scales the maximum permitted displacement ceiling independently for bond lengths, valence angles, and dihedral angles. These ceilings are tied to the current training phase $\phi$, decreasing by an order of magnitude across phases to transition the agent from broad PES exploration to fine, chemically precise displacements in the vicinity of the TS. The coordinate-type-resolved scaling is applied as:

$$\mu = 0.1 + 0.9\frac{a_3 + 1}{2} \in [0.1, 1.0] \tag{7}$$

$$\Delta q_i = \Delta q_{\text{raw},i} \cdot \min\left(1, \frac{\mu\, \delta_k^{(\phi)}}{\max_{j \in \mathcal{C}_k} |\, \Delta q_{\text{raw},j}\, |}\right), i \in \mathcal{C}_k \tag{8}$$

where $\mathcal{C}_k \in \{\mathcal{B}, \mathcal{A}, \mathcal{D}\}$ denotes the index sets of bond lengths $\mathcal{B}$, valence angles $\mathcal{A}$, and dihedral angles $\mathcal{D}$, and $\delta_k^{(\phi)}$ is the phase-dependent ceiling for coordinate type $k$. This per-type normalization prevents any single large dihedral rotation from contaminating the physically smaller bond length displacements, preserving the integrity of the molecular geometry at each step. The convergence criteria applied to determine episode termination are tightened correspondingly across phases, ensuring that the final reported TS satisfies the first-order saddle point criteria. Finally, the scaled internal-coordinate displacement $\Delta q_{raw}$ is converted back to Cartesian coordinates through the iterative back-transformation as implemented in the geomeTRIC package, which completes a given timestep of a trajectory (episode).

## Reward Function

The reward signal $R(s, a)$ received by the agent at each timestep is constructed from several additive components designed to shape behaviour across the distinct geometric regimes encountered during TS search. A continuous progress term rewards the agent for reducing the internal-coordinate distance to the target geometry,

$$r_{\text{progress}} = \left\|\mathrm{d}_{\text{target}}\right\|_{t-1} - \left\|\mathrm{d}_{\text{target}}\right\|_{t} \tag{9}$$

ensuring a dense gradient signal throughout the episode. Simultaneously, the norm of the perpendicular gradient $\|\mathbf{g}_{\perp}\|$ is subtracted as a penalty at every step, encouraging the agent to remain on or near the minimum energy path rather than wandering across the PES.

The dominant shaping of the reward, however, is conditioned on the number of negative eigenvalues $N_{\text{neg}}$ of the internal-coordinate Hessian, which serves as the primary indicator of the local topological character of the PES. Penalties are applied when $N_{\text{neg}} = 0$ or $N_{\text{neg}} > 1$ to steer the agent away from minimum-energy regions and higher-order saddle points toward the chemically meaningful first-order saddle region. Within the $n_{neg} = 1$ region, an additional consistency check compares the filtered eigenvalue count against the raw imaginary mode ($n_{imag} = 1$) count from the electronic structure calculation. If these quantities disagree, the agent receives a penalty to prevent it from exploiting numerical artifacts near the boundary between curvature regimes. When both indicators consistently confirm a single imaginary mode, the agent is rewarded for reductions in the RMS gradient norm below the best value recorded in the current episode, with increases penalized proportionally at a coefficient that grows across training phases to enforce increasingly strict convergence. Upon discovery of a geometrically distinct TS zone, verified by a tolerance check across all Z-matrix internal coordinates, a large sparse bonus is awarded, with a smaller bonus granted for subsequent improvements within the same zone. Finally, the reward structure includes terminal components at episode termination: a large positive reward is granted upon satisfaction of the absolute first-order saddle point convergence criteria ($g_{norm}$< 0.005 Hartree/Bohr), with magnitude scaled by the quality of convergence, while a large negative terminal penalty is applied in the event of an unphysical geometry or electronic structure failure, discouraging the agent from taking steps that destroy the molecular integrity. The complete reward function together with the complete mathematical formulation of the reward function are described in Supplementary Information.

**Soft Actor-Critic (SAC) Algorithm**

The search for TS geometries requires a policymaker that can effectively balance the exploration of the potential energy surface with the exploitation of saddle point regions. To address this, we employ an actor-critic RL framework in which two coupled neural networks are trained simultaneously: an actor that proposes geometric displacements of the molecular internal coordinates given the current PES state, and a critic that evaluates those proposals by estimating the expected long-term reward, effectively guiding the actor on whether a given structural change is likely to lead toward a saddle point or an unphysical geometry. Among actor-critic methods, we specifically adopt the Soft Actor-Critic (SAC) algorithm,[52] an off-policy maximum entropy framework that has demonstrated superior performance on complex continuous control tasks owing to its stability and sample efficiency, for training the agent. The training follows a dual objective where the agent seeks to maximize the expected cumulative reward and the policy entropy ($\mathcal{H}$).

$$J(\pi) = \sum_{t} \mathbb{E}_{(s_t,a_t)\sim\rho_\pi}\left[r(s_t,a_t) + \alpha\mathcal{H}\big(\pi(\cdot\,|s_t)\big)\right] \tag{10}$$

Following the formulation of Haarnoja et al.,[53] the temperature parameter $\alpha$ is treated as a learnable variable updated at each training iteration to satisfy a target entropy constraint. This adaptive mechanism is particularly advantageous for navigating complex molecular landscapes, as it enables the agent to dynamically balance exploration and exploitation. By automating this balance, the agent maintains a consistent search policy without the need for system-specific hyperparameter calibration, thereby enhancing the transferability of the framework across diverse chemical systems. All necessary hyperparameter values and PyTorch implementation details can be found in the Supplementary Information.

## Progressive Rendering

The training curriculum is organized into progressive phases that transition from broad exploration to precise convergence. In the initial phase, the agent is permitted to take large displacements across the surface, with maximum step sizes of 0.05 Bohr for bonds, 0.10 radians for angles, and 0.30 radians for dihedrals, scaled by a learned speed controller ($a_3$) that allows the agent to modulate its pace between 10% and 100% of these limits. The geometric tolerances for reaching the target product are similarly relaxed, requiring only that cumulative bond, angle, and dihedral deviations fall below 0.20, 0.10, and 0.15 respectively. This permissive setting encourages the agent to traverse large portions of the PES within the 200-step episode horizon, prioritizing the discovery of saddle point regions over geometric precision. The transition to the second phase is triggered once the agent discovers a candidate TS zone, defined as a geometry exhibiting exactly one negative Hessian eigenvalue with a gradient norm below 0.01 Hartree/Bohr, and confirmed as geometrically distinct from all previously recorded zones through a tolerance check across the full set of internal coordinates. Rather than requiring the agent to repeatedly climb out of the reactant well in subsequent episodes, the second phase repositions the starting geometry to the last point along the successful trajectory where no negative eigenvalues were present, effectively zooming into the neighbourhood of the discovered saddle region. This eliminates redundant traversal of the reactant basin and redirects the full exploration budget toward resolving the precise stationary point. The maximum displacement per step is reduced by an order of magnitude to 0.005 Bohr for bonds, 0.010 radians for angles, and 0.020 radians for dihedrals, and the geometric convergence tolerances tighten correspondingly to 0.10, 0.05, and 0.10. The gradient penalty coefficient also increases, enforcing stricter descent toward the saddle point than in the exploratory phase with a new convergence criterion of below 0.005 Hartree/Bohr. To ensure the agent retains sufficient room to converge under these tighter constraints, the episode horizon is extended to 400 steps, compensating for the smaller displacements with a longer trajectory budget.

### Computational Details

All electronic structure calculations in this work are performed using the GFN2-xTB semiempirical tight-binding method as implemented in the xtb program package.[54] Internal coordinate transformations are handled using the geomeTRIC[51] optimization library. geomeTRIC constructs a redundant set of primitive internal coordinates consisting of bond distances, bond angles, and dihedral angles from the molecular connectivity, and provides the Wilson B-matrix that maps between internal and Cartesian coordinate spaces. Nudge elastic band (NEB) calculations with improved tangent method were performed using the Atomic Simulation Environment (ASE),[55] with forces evaluated at the same GFN2-xTB level of theory to ensure a consistent comparison with the SACTS agent.

## Results and Discussion

To evaluate SACTS, we tested the model on a series of benchmark reactions,[56, 57] spanning systems of increasing molecular complexity and PES ruggedness. At each timestep, the agent interacts with an environment driven by the GFN2-xTB,[54] with geometric transformations handled by the geomeTRIC package, together providing the energy, gradient, and Hessian information necessary to evaluate the local PES topology and construct the basis vectors governing the agent's geometric displacements. Training progress is monitored through the cumulative episodic return and average reward metrics, and the learning process of SACTS is divided into two discrete phases allowing for coarse-to-fine refinement, with transitions between phases triggered by convergence thresholds in the gradient. We demonstrate the application of the method using the HCN isomerization reaction, a canonical benchmark system with a readily visualizable PES.

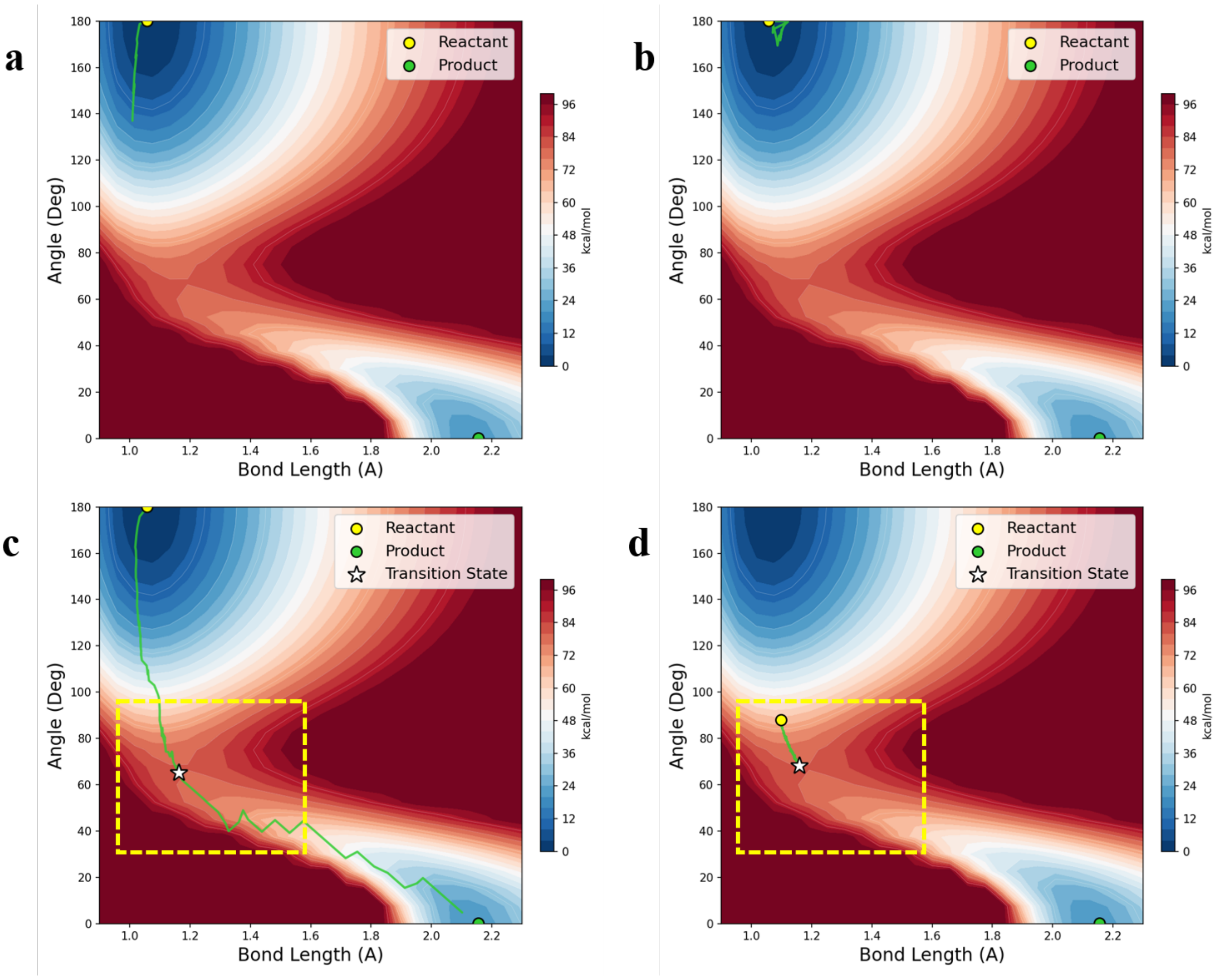

**Figure 2.** Representative trajectories of the SACTS agent on the potential energy surface of the HCN ⇌ HNC isomerization reaction, illustrating the progression from naive exploration to targeted convergence. (a) Phase 0 (broad exploration): Random policy of the agent explores a region of the surface before terminating due to an unphysical geometry. (b) Subsequent episode in which the agent overshoots the accessible region and terminates at the PES boundary (c) The agent successfully navigates from the reactant basin into the first-order saddle region,

satisfying the TS detection criteria and identifying a candidate zone. (d) Phase 1 (progressive rendering, tight refinement): starting from a spawn point near the identified zone, the agent converges monotonically to the TS, satisfying the absolute convergence criteria.

As shown in the Figure 2, the agent initiates its search from the reactant geometry and navigates the PES toward the target product. Rather than relying solely on the softest vibrational mode, which can lead to non-productive pathways and conformational drift, the agent's exploration is guided by the integration of a target vector that provides directional information toward the product geometry. In the earliest episodes, the agent has no knowledge of the PES landscape, and its policy is effectively random making it overshoot bond length boundaries, wandering into high-energy repulsive regions, entering higher-order saddle points where multiple eigenvalues are negative, and frequently trigger episode termination through steric clashes or unphysical geometries. These failures show up clearly in Figure 3, where the agent's missteps into unphysical conformations appear as sharp negative spikes in the reward that cut short the early, darker coloured episodes. The failed trajectories, however, are not wasted: the large negative penalties received at the boundaries and the growing time penalties accumulated in minimum-energy basins propagate through the replay buffer into the value function, teaching the agent which regions of configuration space to avoid and, critically, which directions lead away from them. By the later episodes, the agent exhibits behaviour that resembles the iterative refinement of a conventional TS optimizer: it approaches the saddle region along the softest Hessian mode, reduces the perpendicular gradient monotonically, and converges to a structure satisfying the first-order saddle point criteria.

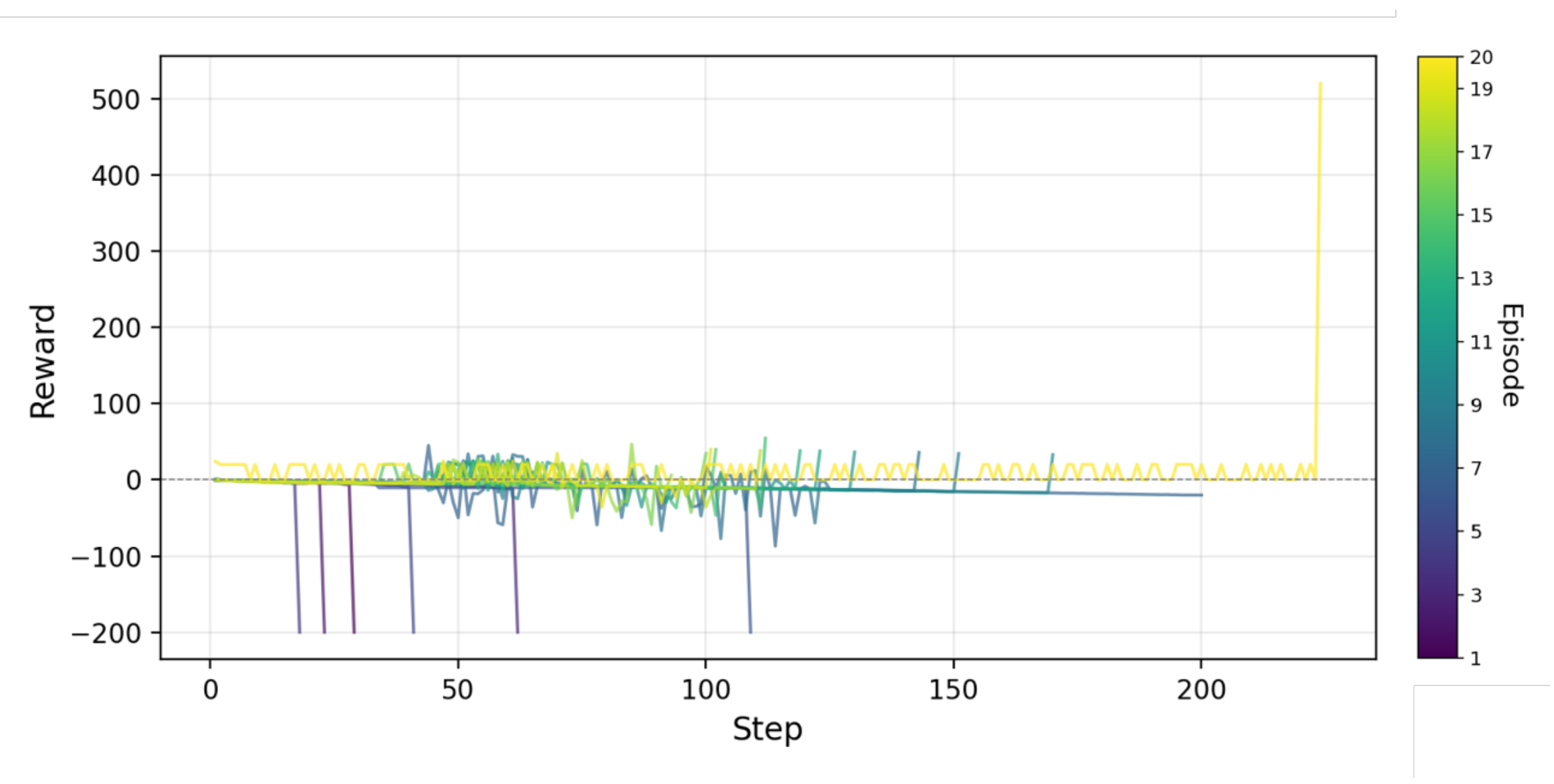

**Figure 3.** Reward per step for the HCN isomerization TS search with dark-to-light colour gradient maps progression from early to late episodes.

Upon satisfying the convergence criteria of phase 0, the agent transitions to phase 1 (progressive rendering, Figure 2(d)), spawning from the last minimum-energy geometry recorded prior to entering the saddle region. This progressive refinement strategy confines the search to the vicinity of the best TS structure identified in the preceding phase, with the agent now operating under tighter step-size constraints to yield a precise saddle-point geometry. For the representative system shown in Figure 3, phase 0 required 19 episodes averaging approximately 100 steps each out of the 200 step threshold to satisfy its transition criteria, while phase 1 converged to the absolute saddle-point threshold of 0.005 Hartree/Bohr within a single episode, illustrating the efficiency gained by repositioning the agent directly into the saddle region. The phase transition and the accompanying improvement in convergence quality are reflected as an abrupt increase in the episodic reward visible in the training curve (Figure 3). A notable feature of the training curve during phase zero is the intermittent dips in episodic reward that persist even as the agent approaches convergence. These arise from the maximum-entropy objective of the SAC algorithm, which augments the expected return with an entropy bonus on the policy distribution, preventing premature collapse onto a single trajectory. Rather than exploiting the current best path exclusively, the agent continues to sample alternative step sequences in the vicinity of the saddle, effectively probing whether neighbouring regions of the PES contain a lower-gradient stationary point or a geometrically distinct TS. This entropy-driven exploration serves as an implicit safeguard against convergence to a suboptimal saddle point. The agent terminates its search either upon reaching absolute convergence criteria or by a vanishing gradient with exactly one negative Hessian eigenvalue or upon exceeding the maximum episode limit. All algorithmic hyperparameters and implementation details are summarized in the Supplementary Information.

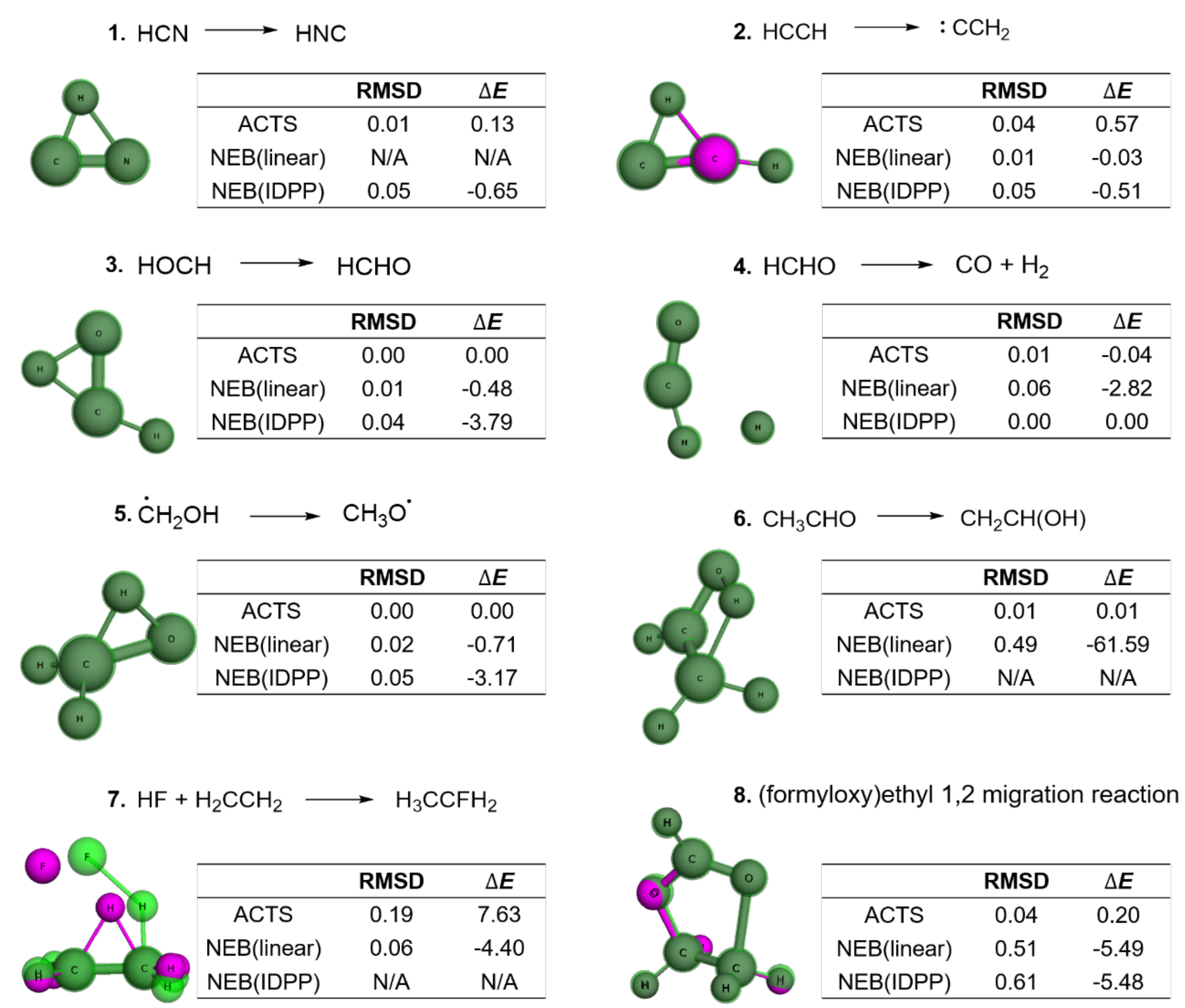

**Figure 4**. RMSD (in Å) and energy differences (in kcal mol$^{-1}$) of TSs located by SACTS and NEB with respect to reference structures optimized via the Berny algorithm. The TSs obtained by SACTS is shown with pink color, while the reference geometry with green color.

The SACTS performance on a small set of test reactions with an increasing number of atoms and TS complexity is summarized in Figure 4. All SACTS-generated geometries are compared against TS located with the Berny optimization algorithm. To place these results in context, we also performed NEB calculations with the improved tangent formulation using both linear and image dependent pair potential (IDPP)[58] interpolation schemes across all eight reactions. SACTS recovers the saddle point geometry across all eight benchmark systems with RMSD values consistently at or below 0.04 Å and energy differences staying well under 1 kcal mol$^{-1}$with the only exception being the addition of HF to ethene (reaction **7**). The smallest deviations are found for HOCH → HCHO and ·$CH_2OH$ → $CH_3O$·, where both RMSD and energy difference are in agreement with the Berny TS algorithm, consistent with the simple topology of these rearrangements. NEB with linear interpolation performs comparably on the smaller systems (reactions **2** through **5**), achieving low RMSDs though with somewhat larger energy deviations, while NEB with IDPP interpolation locates reasonable geometries for reactions **1** through **5** but consistently yields larger energy differences than either SACTS or NEB with linear interpolation on these same systems. The contrast between the methods sharpens considerably for the more complex reactions. For reaction **6**, the NEB-linear converges to the product structure rather than locating the saddle point, yielding an RMSD of 0.49 Å and Δ*E* of -61.59 kcal mol$^{-1}$ against the Berny reference and the NEB-IDPP fails to converge entirely for this system. SACTS, by contrast, locates the saddle point with an RMSD of 0.01 Å and a ΔE of just 0.01 kcal mol$^{-1}$. For the HF addition to ethene (reaction **7**), NEB with linear interpolation achieves a low RMSD of 0.06 Å but with a ΔE of -4.40 kcal mol$^{-1}$, while NEB with IDPP again fails to converge. SACTS reaches an RMSD of 0.19 Å and a ΔE of 7.63 kcal mol$^{-1}$ for this reaction, representing its largest deviation across the benchmark. Prior studies[59] have reported success on comparable systems using climbing image NEB (CI-NEB) combined with internal coordinate based interpolation, suggesting that the choice of interpolation scheme plays a critical role in NEB convergence for complex reaction pathways. Across the full benchmark, SACTS gives the closer geometric and energetic agreement, and in cases where absolute convergence is not reached within the training epoch its refined structures still fall inside the basin of the true TS, making them strong initial guesses for further refinement with conventional optimizers.

The larger deviation seen for the HF addition (reaction **7**) and the use of a masked target vector points to an inherent challenge of the current framework. In reactions that involve substantial geometric reorganization between the reactant and the product, the target displacement vector can grow disproportionately large relative to the contributions from the soft mode and the steering directions, which overwhelms the curvature informed part of the step and introduces noise into the search trajectory. This imbalance can show up as unphysical geometries or as convergence toward saddle points that sit close to the target vector but are not chemically relevant to the reaction of interest. More broadly, the structural awareness of the agent is largely implicit, drawn from PES feedback in the form of energies, gradients, and Hessian curvatures and supported only by simple safety signals such as proximity-based clash detection and bond length bounds. The agent therefore has no explicit knowledge of molecular connectivity or chemical bonding, which limits its ability to tell early TSs from late ones or to recognize when the located saddle point belongs to a competing pathway rather than the intended one. These challenges also point to clear opportunities for future work. Extending the agent to larger reaction networks will call for explicit topological state information and reward functions that more directly guide the search toward the relevant pathway. Studying how learned parameters transfer across different potential energy surfaces could let the agent generalize its strategy, improving efficiency and reducing the effort needed to locate the optimal reaction path.

## Conclusions

We presented a novel RL agent that can autonomously navigate molecular potential energy surfaces and identify first-order saddle points across a chemically diverse set of benchmark reactions. Rather than learning from a static dataset of pre-optimized structures, SACTS acquires its search strategy through direct exploration, guided at each step by curvature, gradient, and structural feedback that together capture a physically meaningful picture of the local PES topology. Furthermore, progressive rendering ensures that this exploration is efficient as beginning with coarse, wide-ranging steps that map out the saddle-point landscape and progressively tightening into a precise refinement once a promising region is identified.

The benchmark results confirm that the geometries produced by SACTS are consistently within chemical accuracy of reference saddle points obtained by established optimization algorithms, and in cases where strict convergence is not reached, the structures are sufficiently close to the true saddle point to serve as reliable starting points for downstream refinement. Demonstrated here on systems of up to 10 atoms with a new agent trained for each reaction, the framework establishes a clear and dependable baseline and lays the groundwork for the extensions that follow. We anticipate that future extensions of this framework will focus on achieving generalized transferability across different potential energy surfaces and scaling the architecture to handle larger, more complex molecular systems, further establishing reinforcement learning as a practical tool for automated reaction discovery.

## Acknowledgements

The authors gratefully acknowledge the CAREER and CAS-Climate programs of the National Science Foundation for financial support of this work (Grant no. CHE-2143354). The authors acknowledge the Infrastructure for Scientific Applications and Advanced Computing (ISAAC) of the University of Tennessee for computational resources.

## Data Availability Statement

The data for the following information are included in the ESI: SACTS algorithm overview, reward function, phase-dependent parameters, neural network architecture and hyperparameters. The code developed for the project that is presented in the submitted paper with the title “Autonomous Transition State Search with Soft Actor-Critic Reinforcement Learning” by Suresh and Vogiatzis is deposited on GitHub (https://github.com/usuresh2509/SACTS.git).

## Conflicts of Interest

The authors declare no competing interests.